# Ultrabroadband Density of States of Amorphous In-Ga-Zn-O

Kyle T. Vogt,[1,*] Christopher E. Malmberg,[2] Jacob C. Buchanan,[2] George W. Mattson,[1] G. Mirek Brandt,[1,†] Dylan B. Fast,[2] Paul H.-Y. Cheong,[2,*] John F. Wager,[3,*] Matt W. Graham[1,*]

[1.] Department of Physics, Oregon State University, Corvallis, OR 97331-6507 USA
[2.] Department of Chemistry, Oregon State University, Corvallis, OR 97331-4003 USA
[3.] School of EECS, Oregon State University, Corvallis, OR 97331-5501 USA

**ABSTRACT:** The sub-gap density of states of amorphous indium gallium zinc oxide ($a$-IGZO) is obtained using the ultrabroadband photoconduction (UBPC) response of thin-film transistors (TFTs). Density functional theory simulations classify the origin of the measured sub-gap density of states peaks as a series of donor-like oxygen vacancy states and acceptor-like Zn vacancy states. Donor peaks are found both near the conduction band and deep in the sub-gap, with peak densities of $10^{17}$-$10^{18}$ cm$^{-3}$eV$^{-1}$. Two deep acceptor-like peaks lie adjacent to the valance band Urbach tail region at 2.0 to 2.5 eV below the conduction band edge, with peak densities in the range of $10^{18}$ cm$^{-3}$eV$^{-1}$. By applying detailed charge balance, we show that increasing the deep acceptor density strongly shifts the $a$-IGZO TFT threshold voltage to more positive values. Photoionization ($hv > 2.0$ eV) of deep acceptors is one cause of transfer curve hysteresis in $a$-IGZO TFTs, owing to longer recombination lifetimes as electrons are captured into acceptor-like vacancies.

**INTRODUCTION:** Oxide thin-film transistors (TFTs) have gained traction as driver components for active-matrix liquid crystal and organic LED displays as pixel control circuits.[1–4] Amorphous In-Ga-Zn-O ($a$-IGZO) in particular has become a successful alternative to amorphous silicon for manufacturing TFTs with high mobility[5] and low leakage current,[6] enabling large-area display applications. In $a$-IGZO, subtle variations in composition or processing create sub-gap defect and vacancy states[7–12] that control both TFT semiconducting behavior[13] and performance limitations.[14] The ability to measure and identify the structural origins of these sub-gap states is crucial to understanding the electrical behavior of *a*-IGZO TFTs. Unfortunately, the disordered nature of amorphous oxide thin films such as *a*-IGZO makes this an unsolved problem. In addition, measuring sub-gap trap density in amorphous thin film materials like IGZO presents many inherent challenges. Not only is the sub-gap state concentration small ($< 10^{18}$ cm$^{-3}$), but the TFT threshold voltage tends to drift, making sensitive transport and photoconductive measurements challenging.

Substantial work has been reported on the characterization of defect states in *a*-IGZO from first principles,[15–18] and experiments utilizing both photoexcitation[19–21] and electrical[22,23] methods. Most notably, lamp-based optical illumination methods performed on *a*-IGZO TFTs, such as photoexcitation charge collection spectroscopy (PECCS),[24] have been successful in quantifying the density of defect states in the sub-gap. TFT behavior under illumination has also been shown to depend strongly on photon energy ($h\nu$), especially upon photoexcitation of "deep states" near the valence band,[25–28] which suggests the existence of multiple species of sub-gap states. However, the near-bandgap photoexcited TFT behavior has been multiply attributed to defects related to hydrogen,[29–31] excess oxygen,[32–34] and contradictorily, the lack of oxygen.[35] Recently, Jia *et al.* suggested the existence of cation vacancy ($V_M$)-related clusters in IGZO as a vehicle for the inclusion of stable excess oxygen.[36] Thus, ambiguity still remains about the exact structural origin of sub-gap states.

In this work, eight Gaussian-like sub-gap states are observed in *a*-IGZO by measuring the photoconduction response of a TFT over a photon range of 0.3 to 3.5 eV using tunable lasers. Spectrally resolved photoconduction decay lifetime and TFT drain current-gate voltage transfer curve hysteresis experiments suggest that six of these peaks are donor-like sub-gap states, while the other two peaks are acceptor-like states. Ab-initio DFT+U simulations of vacancy defects attribute the donor-like peaks to oxygen vacancies, whereas the acceptor-like peaks are likely attributed to Zinc vacancies. Zinc vacancy identification is accomplished by calculating the formation energy of candidate acceptor-like point defects, i.e., indium, gallium, or zinc vacancies or oxygen self-interstitials on oxygen-poor *a*-IGZO. When the Fermi level is positioned near the conduction band edge, the zinc vacancy is found to have the smallest formation energy of the acceptor-like defects considered. Moreover, the zinc vacancy formation energy is found to be negative such that zinc vacancies appear to be generated as a consequence of self-compensation; the self-compensation mechanism resolves the puzzle of why a metal vacancy is formed in a material such as a-IGZO that is known to be anion-deficient.

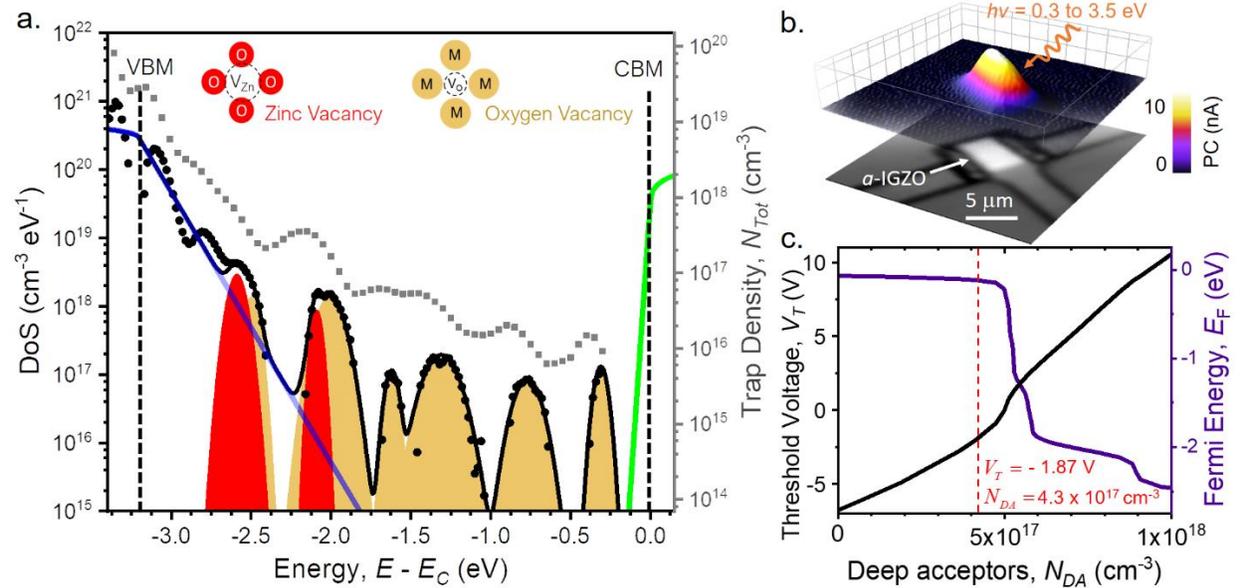

**Figure 1 (a)** The total integrated sub-gap trap density ($N_{Tot}$, *gray squares*) of *a*-IGZO, as measured by ultrabroadband photoconduction (UBPC). The density of states (DoS, black circles) is the first derivative of the gray curve. Sub-gap DoS peaks originate from oxygen vacancies (*gold peaks*) and Zn vacancies (*red peaks*). The green (*blue*) line is the conduction (valence) band and band tail states. **(b)** The *a*-IGZO TFT photoconduction response (PC, color bar) is proportional to $N_{Tot}$, overlaid with TFT reflectance map (*grayscale*). **(c)** Using the UBPC-derived DoS, we invoke charge balance to extrapolate both the TFT $V_T$ and $E_F$ dependence on to the metal vacancy deep acceptor concentration ($N_{DA}$). The red line corresponds to the total integrated concentration of Zn vacancy states measured for the *a*-IGZO TFT shown above.

**RESULTS:**

Figure 1a plots a representative *a*-IGZO DoS profile as determined by our ultrabroadband photoconduction technique (UBPC; see Experiment and Analysis for a detailed discussion of the UBPC method). The DoS profile in Fig. 1a derives from the *a*-IGZO TFT device photoconduction (PC) response, shown in Fig. 1b (color map), which is collected for laser excitation energies ranging from $h\nu$ = 0.3 to 3.5 eV. The resulting PC spectrum is proportional to the total sub-gap trap density, $N_{Tot}$ (cm$^{-3}$) plotted in Fig. 1a (gray squares). The sub-gap DoS ($g(E - E_c)$) in Fig. 1a (black circles) is obtained by numerical differentiation with respect to energy of $N_{Tot}$, i.e., $g(E - E_c) = dN_{Tot}/dE$ (cm$^{-3}$ eV$^{-1}$), where energy is given in reference to the conduction band minimum (CBM), i.e., $E_C = 0$. In addition, both the band tail state density and near band edge states are plotted for both the CBM[37] (green line) and the valence band maximum (VBM) region (blue line, this work).

Eight sub-gap peaks are identified in Fig.1a and are fitted with Gaussian distributions. Six of these peaks are ascribed to oxygen vacancies, $V_O$ (gold shading), while the two other peaks are attributed to Zn vacancies, $V_{Zn}$ (red shading). $V_O$ are donors, i.e., neutral when filled with two electrons, and positively charged (2+ ionized) when left unfilled, whereas $V_{Zn}$ are acceptors, i.e., negatively charged (2- ionized) when filled with electrons and neutral when left unfilled. A detailed discussion on defect peak identification will follow later using DFT (see First-Principles DoS Analysis), charge balance considerations, and electron capture lifetimes measurements.

The peak energy ($E_T$), density ($D_T$), and width ($w$, full 1/$e$ width) of Gaussian sub-gap state distributions as well as valence band tail parameters, obtained from the DoS in Fig. 1a are summarized in Table 1. The sum of these distributions constitutes the black line in Fig. 1a. The sub-gap state densities and VB Urbach tail parameters in Table 1 are similar to values reported from various experimental techniques.[7,21,38–43] These previous measurements are typically limited to characterizing either shallow states near the CBM or deep states near the VBM; whereas the DoS profile obtained from UBPC extends continuously from 0.3 to 3.5 eV. Moreover, the assignment acceptor-like states lying deep in the sub-gap is typically considered unimportant since holes forming in these states are not mobile, although they are thought to play an important role in negative bias illumination stressing (NBIS).[44]

| DoS Peak | $E_T$ (eV) | $D_T \times 10^{17}$ (cm$^{-3}$ eV$^{-1}$) | $w$ (eV) |
|---|---|---|---|
| **O-1** | -0.31 | 1.31 | 0.10 |
| **O-2** | -0.77 | 0.85 | 0.20 |
| **O-3** | -1.31 | 1.95 | 0.25 |
| **O-4** | -1.62 | 1.15 | 0.10 |
| **O-5** | -2.01 | 13.9 | 0.20 |
| **O-6** | -2.50 | 13.0 | 0.11 |
| **Zn-7** | -2.09 | 8.45 | 0.08 |
| **Zn-8** | -2.59 | 30.0 | 0.14 |
| **VB Tail** | $E_g$=3.12 | 2800 | 0.111 |

Table 1: *a*-IGZO DoS figures of merit. Peaks are broken in to two categories, **O** and **Zn**, which indicate the origin of the defect as being either from oxygen or zinc vacancies.

TFT performance characteristics such as carrier mobility are typically associated with near CB defect states, which could lead one to suspect that deep sub-gap states play no role in determining the performance of an *a*-IGZO TFT. We now apply the experimentally measured DoS to present two situations in which Zn vacancy deep acceptors play an important role in affecting *a*-IGZO TFT device behavior.

The first example involves calculating the Fermi energy ($E_F$) and threshold voltage ($V_T$) as a function of the total Zn vacancy deep acceptor concentration ($N_{DA}$). As shown in Fig. 1c, both the TFT $V_T$ and $E_F$ depend critically on $N_{DA}$. This observation is a consequence of charge balance in *a*-IGZO, given by

$$N_{SD} + N_{DD}^{+}(E_F) + N_{TD}^{+}(E_F) = N_{DA}^{-}(E_F) + n(E_F) + N_{TA}^{-}(E_F), \qquad (1.1)$$

where $N_{SD}$ is the total density of shallow V$_O$ donors, $N_{DD}^{+}(E_F)$ is the density of ionized deep V$_O$ donors, $N_{TD}^{+}(E_F)$ is the density of ionized valence band tail donor states, $N_{DA}^{-}(E_F)$ is the density of ionized deep V$_{Zn}$ acceptors, $n(E_F)$ is the free electron density in the conduction band, and $N_{TA}^{-}(E_F)$ is the density of ionized conduction band tail acceptor states. Note that $N_{DD}^{+}$, $N_{TD}^{+}$, $N_{DA}^{-}$, $n$, and $N_{TA}^{-}$ are explicitly shown to be dependent on $E_F$, which can be evaluated by equating positive and negative charge, as specified in Eq. 1.1. UBPC DoS measurements in Fig. 1a. The above charge balance considerations suggest that, $N_{DA} \approx 4.3 \times 10^{17}$ cm$^{-3}$, and $N_{SD} \approx 5 \times 10^{17}$ cm$^{-3}$ which results in a $V_T \approx -1.87$ V (red dotted line in Fig. 1c, and confirmed by experimental TFT transfer curve in the Supplemental Information). The threshold voltage, $V_T$, is estimated using the discrete donor trap model[45] once $E_F$ is known, given by,

$$V_T(E_F) = \frac{q}{C_I}\left[\left(N_{TD}^+\right)^{2/3} + \left(N_{DD}^+\right)^{2/3} + \left(N_{DA}^0\right)^{2/3} - (n)^{2/3}\right] \qquad (1.2)$$

where $q$ is the electron charge, $C_I$ is the insulator capacitance density ($\sim$ 11.5 nF cm$^{-2}$, for 300 nm of SiO$_2$ gate dielectric), and $N_{DA}^0$ is the density of neutral Zn vacancy deep acceptor states.

As V$_{Zn}$ deep acceptor density increases from $10^{17}$ to $10^{18}$ cm$^{-3}$ in Fig. 1c, $V_T$ increases monotonically from $\sim$ –5 V to $\sim$ 10 V. $V_T$ is negative (depletion mode) when $N_{DA} < N_{SD}$, and positive (enhancement mode) when $N_{DA} > N_{SD}$. This trend in $V_T(N_{DA})$ can be understood further by examining the $E_F(N_{DA})$ curve included in Fig. 1c. If $N_{DA} < N_{SD}$, $E_F$ is positioned very close to the conduction band minimum so that $n$ becomes the dominant negative charge term in Eq. 1.1, resulting in the depletion-mode TFT behavior due to the formation of an accumulation layer. For this limiting range of $N_{DA}$, $E_F$ remains almost constant in Fig. 1c. However, this trend in $E_F$ changes abruptly when $N_{DA} \approx N_{SD}$, as $E_F$ drops precipitously to $\sim$ 1.2 eV below the CBM. This abrupt change in $E_F$ (and concomitant fast increase of $V_T$) is a consequence of the very small ($\sim 1 \times 10^{16}$ cm$^{-3}$) densities of both the O-1 and O-2 sub-gap states; as $N_{DA}$ is increased, both of these states are easily emptied (and ionized) as $E_F$ pushes towards the valence band in order to maintain charge balance. A second less abrupt drop in $E_F$ occurs as the higher density O-3 peak ($n_{O-3} = 4.3 \times 10^{16}$ cm$^{-3}$) is ionized, until it encounters the much higher densities of the O-5 ($n_{O-5} = 2.5 \times 10^{17}$ cm$^{-3}$) and the Zn-7 ($n_{M-7} = 6.0 \times 10^{16}$ cm$^{-3}$) peaks, which tend to clamp the $E_F$ near ~ 2 eV below the conduction band minimum. A final step terminates at ~ 2.5 eV when $E_F$ encounters O-6 ($n_{O-6} = 1.3 \times 10^{17}$ cm$^{-3}$), Zn-8 ($n_{M-8} = 3.7 \times 10^{17}$ cm$^{-3}$) as well as valence band tail states. This $E_F(N_{DA})$ trend reveals that strong enhancement-mode TFT behavior, i.e., $V_T$ ~ 10 V, is associated with the existence of empty V$_O$, V$_{Zn}$, and valence band tail states in the sub-gap, as a result of charge balance, given Zn and oxygen vacancy DoS throughout the sub-gap.

Our second example shows how Zn vacancy deep acceptors impact *a*-IGZO TFT performance under illumination by tuning photoexcitation energies in order to preferentially excite either V$_O$ or V$_{Zn}$ type vacancies. Figure 2a shows a comparison of a drain current-gate voltage ($I_D$-$V_G$) transfer curve for an *a*-IGZO TFT illuminated with $h\nu$ =1.4 eV (Fig. 2a, gold curve) or at 2.5 eV (Fig. 2a, red curve). While the transfer curve at 1.4 eV is almost identical to the curve measured under dark conditions, the $h\nu$ = 2.5 eV curve deviates greatly. In particular, the turn-on of the $h\nu$ = 1.4 eV illuminated curve is well behaved, showing almost no hysteresis. By contrast, the $h\nu$ = 2.5 eV curve is highly non-ideal, exhibiting a large amount of clockwise hysteresis and a much higher $V_T$ shift, with the maximum current decreased, and the minimum current increased. The hysteresis voltage, V$_H$, is given as the separation of forward and reverse sweeps at $I_D$ = 10$^{-10}$ A.

Figure 2b illustrates another photoinduced trend, in which the PC response of an *a*-IGZO TFT is monitored during and after laser excitation (see inset of Fig. 2b). After $h\nu$ = 1.4 eV photoexcitation (Fig. 2b, gold curve) the fall time is noticeably shorter than the fall time associated with $h\nu$ = 2.5 eV photoexcitation (Fig. 2b, red curve). To quantify this, the average PC decay fall time (τ) is given as the 1/*e* lifetime of the longest monoexponential component (dashed line fits), which is attributed to electron recombination with vacancy defect sites.

Hysteresis and PC-lifetime trends as a function of photon energy are displayed in the upper two panels of Fig. 2c. Comparison of Fig. 2ci and ii shows a clear onset exists when $h\nu > \sim$ 2 eV, above which the hysteresis voltage and the PC fall time both abruptly increase. We ascribe this onset to the initiation of Zn vacancy photoexcitation. The key point here is that recombination of photoexcited electrons is found to be more sluggish when electrons are photoexcited from $V_{Zn}$ vacancy states than from oxygen vacancy states. Why is this?

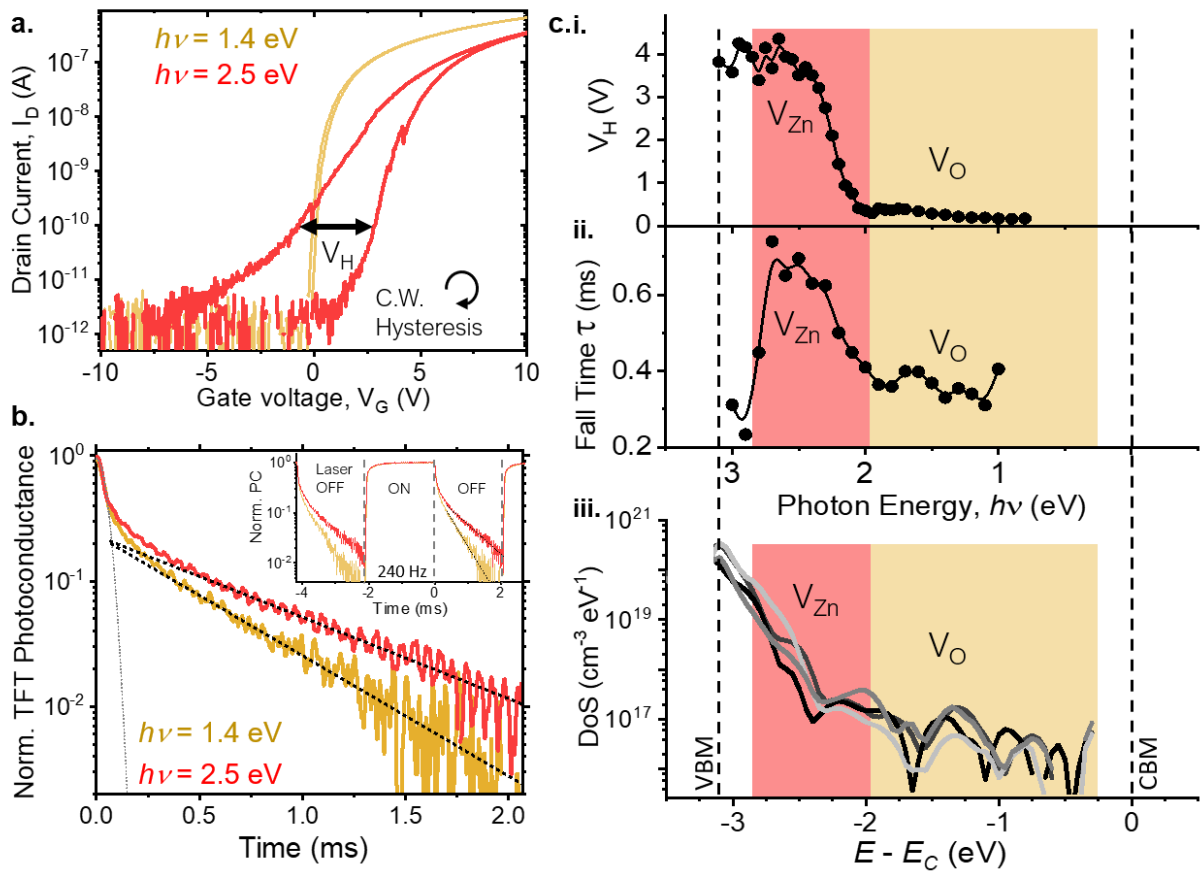

**Figure 2 (a)** Two TFT $I_D$-$V_G$ transfer curves differ starkly with the photon energy of illumination. **(b)** Transient PC decay curves for the same two photon energies. Black dashed lines are exponential fits to the longest decay component (fall time, τ). The gray dotted line gives the instrument response. **(c) i.** Hysteresis voltage, $V_H$, versus photon energy. **ii.** Photoconduction decay fall time, τ, versus photon energy. **iii.** DoS plots from four TFTs fabricated using different processing conditions. Regions dominated by $V_{Zn}$ ($V_O$) are shaded *red* (*gold*).

We believe that this photoexcited electron recombination trend is associated with the acceptor- or donor-like nature of the Zn or oxygen vacancy, respectively. This follows from the fact that the electron capture time, $\tau_c$, is given by $\tau_c \approx (\sigma_n v_{th} n)^{-1}$, where $\sigma_n$ is the electron capture cross section, $v_{th}$ is the electron thermal velocity (~ $10^7$ cm s$^{-1}$), and $n$ is the free electron concentration. Electron capture by a donor-like defect (e.g., an oxygen vacancy in $a$-IGZO) is a coulombically-attractive process, i.e., an empty donor is positively charged such that a recombining electron experiences coulombic attraction prior to capture, thus enhancing the probability of its capture. In contrast, electron capture by an acceptor-like defect is an electrostatically neutral process, i.e., an empty acceptor is neutral such that there is no coulombic enhancement to electron capture. The capture cross section for neutral capture is given as $\sigma_n$ ~ $10^{-15}$ cm$^2$ and for coulombically-attractive capture is $\sigma_n$ ~ $10^{-12}$ cm$^2$,[46] implying that electron capture to donor states is intrinsically more rapid than electron capture to acceptor states. (Note that the capture cross section for coulombic-repulsive capture, as relevant to the capture of a second electron in a zinc vacancy, is expected to be much smaller than that of neutral capture.) Reexamining Fig. 2b and Fig 2cii, it becomes clear why the PC decay time is maximal near $h\nu$ ~ 2.5 eV, since this corresponds to the only region of the spectrum where there are a majority of Zn vacancy defect states. This agrees well with predictions based on capture cross section.

Using the simple expression for the electron capture time, $\tau_c$ and the electron concentration, $n$, we now assess the plausibility of the measured $a$-IGZO defect electron capture time scales. When an $a$-IGZO TFT is turned on, the accumulation layer electron concentration can easily reach a value of ~ $10^{18}$ cm$^{-3}$ (or larger), such that electron recombination is extraordinarily rapid, e.g., $\tau_c$ ~ $10^{-10}$ s (acceptor) or ~ $10^{-13}$ s (donor). However, far away from the accumulation layer where the semiconductor is in depletion, the electron concentration can be $10^8$ cm$^{-3}$ (or smaller), such that electron recombination is much slower, e.g., $\tau_c$ ~ 1 s (acceptor) or

~ $10^{-3}$ s (donor). Thus, since $n$ spans many orders of magnitude over the thickness an $a$-IGZO TFT structure, an extremely wide range of electron capture times is accessed in a photoexcitation experiment. As a consequence, we argue that the measured photoexcitation fall times of $\tau \sim 400$ μs and ~ 700 μs for donor- and acceptor-like trap states, respectively, are reasonable since each measured time constant constitutes a spatial average of all possible recombination times for electrons throughout the $a$-IGZO layer.

The photoconduction transient fall times involve a time scale much different than that of $I_D$-$V_G$ transfer curve hysteresis, i.e., ~ $10^{-4}$ – $10^{-3}$ s, and ~10 s, respectively. Thus, even though each phenomenon is attributed to Zn vacancy photoexcitation, the dynamics of these processes will differ significantly since the PC decay curves are measured when the TFT is in accumulation ($V_G \gg V_T$), while the hysteresis measurements will require the TFT gate voltage to be held at a strong negative bias for part of the $I_D$ - $V_G$ scan, severely limiting the number free carriers in the device while simultaneously exciting trap states. Nonetheless, both of these measurements are concerning the charges which are able to be quickly recovered after photoexcitation, as opposed to the extremely long recovery of charges (> $10^4$ s) often reported in NBIS experiments.[47]

Finally, Fig. 2ciii plots the DoS of four high-quality $a$-IGZO TFTs fabricated with different processing conditions. While the DoS curves exhibit some variability (less than one order of magnitude), the overall similarity of curves in Fig. 2ciii, suggests the DoS profiles presented here are characteristic of most high-quality $a$-IGZO TFTs. Comparison of Fig. 2ciii to its upper panels might lead the reader to think that the abrupt increase in $V_T$ and $\tau$ could be associated with the onset of the Urbach tail states. However, note that Fig. 2ciii is plotted on a log scale, and the Urbach tail increases exponentially to the valence band edge, whereas $V_T$ and $\tau$ (plotted on a linear scale) only increase until around ~2.6 eV, making their association with Urbach tail states unlikely. Therefore, taken with the above arguments for electron recombination to donor- vs. acceptor-like states, the red shaded area in Fig. 2ciii most likely contains the extent of Zn vacancies in the DoS of $a$-IGZO, which are responsible for the strong TFT dependencies on photoexcitation when $hv$ > 2.0 eV.

## EXPERIMENTAL TECHNIQUE AND ANALYSIS:

The UBPC experimental setup, shown in Figure 3a, chiefly consists of multiple tunable laser sources coupled into a modified scanning photocurrent microscope[48] (SPCM). Light coupling is realized using all-reflective optics, enabling diffraction-limited excitation of $a$-IGZO TFTs with tunable laser sources that continuously span a photon energy range of $hv$ = 0.3 to 3.5 eV (see Supplemental Information). The UBPC response was resolved spectrally by isolating the photoinduced current for each laser energy through lock-in amplification. An oscilloscope triggered by an optical chopper captured the PC decay fall times.

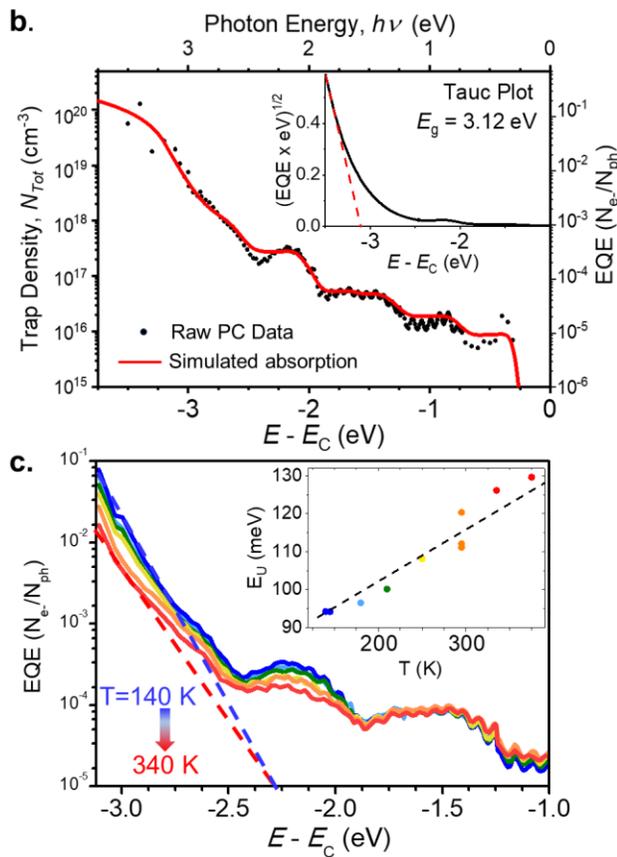

**Figure 3 (a)** Ultrabroadband photoconduction (UBPC) setup. The photoinduced source-drain current is analyzed via **i.** lock-in amplification, **ii.** $I_D$ - $V_G$ curves, or **iii.** an oscilloscope. **(b)** Integrated trap density (*left axis*) and EQE (*right*) vs energy in reference to CBM and photon energy. Simulated absorption (*red*). *(inset)* Photoconductive Tauc plot showing a 3.12 eV bandgap. **(c)** Temperature dependence of UBPC EQE spectra. *(inset)* VB Urbach energy increases with temperature.

To find the total sub-gap trap density, $N_{Tot}$, the PC spectrum is first collected, and normalized by the number of photons, $N_{ph}$, incident on the *a*-IGZO layer at each excitation wavelength. A photon normalized PC spectrum is presented in Figure 3b as the external quantum efficiency (EQE, right axis), given as $\Phi(h\nu) = N_e / N_{ph}$, where $N_e$ is the number of electrons collected due to photoexcitation at the TFT drain. EQE is directly proportional to $N_{Tot}$ (Figure 3b, left axis), with

$$N_{Tot}(h\nu) = \Phi(h\nu) \frac{C_{ox}}{d} \left(\frac{\partial I_D}{\partial V_G}\right)^{-1} \frac{N_{ph}^{max}}{s} \qquad (1.3)$$

where $d$ is the thickness of the *a*-IGZO layer, $C_{ox}$ is the capacitance of the gate insulator (GI); the partial derivative, $\partial I_D / \partial V_G$, is the slope of the TFT (dark) transfer curve, evaluated at the gate voltage used during PC spectral measurements; $N_{ph}^{max} / s$ is the maximum number of photons/s incident on the TFT active area that produces a meaningful increase in PC magnitude[49] and is used to find the absolute trap density. $N_{ph}^{max}$ is found by measuring the incident power at which the PC saturates when photoexcitation energy, $h\nu \sim 3.0$ eV.

Equation 1.3 is a consequence of treating the PC signal as the result of the photofield effect,[50] which causes a shift in the TFT threshold voltage, $\Delta V_T$, when illuminated.[24] The approximation $\Delta V_T(h\nu) \cong \Delta I_D \left(\partial I_D / \partial V_G\right)^{-1}$ is used to extract the shift in $V_T$ due to laser excitation by monitoring drain current, where $\Delta I_D$ is the magnitude of the PC signal, i.e., $\Delta I_D = I_{Light} - I_{Dark}$. $\Delta I_D$ is extracted from the current pre-amp signal (SIG) by the lock-in amplifier. The laser illumination of the TFT is modulated with a mechanical chopper at $f_{chop} \approx 100$ Hz, which serves as the reference (REF) frequency for the lock-in. Lock-in amplification has the essential benefit of isolating the PC contribution from charges that recombine quickly after photoexcitation (e.g., within ~ 10 ms), as well as excluding any slow $V_T$ drift associated with positive bias (or illumination) stress. In this way UBPC is similar to a photon-normalized PECCS[24] spectrum, where the background drift of the threshold voltage is removed via lock-in amplification, and $\Delta I_D$ is monitored in place of $\Delta V_T$ in order to calculate $N_{Tot}$.

The density of states is recovered by differentiating $N_{Tot}$ with respect to energy after suppressing small oscillations associated with thin-film interference using a local numerical regression filter (Loess filter). Figure 3b shows the raw data, while the filtered data was shown as the gray dots in Fig. 1a. Both $N_{Tot}$, calculated as described by Eqn. 1.3, and its derivative (i.e., the DoS) are shown in Fig. 1a. Thus, we establish that the photon normalized PC spectral response is directly proportional to the total trap density, and the DoS is found by taking the derivative of the trap density.

The bandgap, $E_g$, is determined to be ~ 3.12 eV by constructing a Tauc plot[51] from the measured UBPC data (Figure 3b, inset). This is justified since the EQE spectrum is shown to be proportional to the joint density of states given by[52]

$$\alpha(E) \propto C \int_{-\infty}^{\infty} \left[ g(\varepsilon) f_D(\varepsilon - E_{QF}) \right] \left[ g(E-\varepsilon)(1 - f_D(\varepsilon - E_{QF})) \right] d\varepsilon \qquad (1.4)$$

where $C$ is a constant proportional to the coupling of initial and final states (assumed to be independent of photon energy for sub-gap states), $f_D$ is the Fermi-Dirac distribution, $E_{QF}$ is the quasi-Fermi level due to gate bias, and $g(\varepsilon)$ is the DoS in Fig. 1a. Note that $E_{QF}$ is < ~100 meV from the conduction band since all EQE measurements are taken with the device in the "ON" state, i.e., $V_G \gg V_T$. The resulting output of Eq. 1.4, shown in Figure 3b as the red line, agrees remarkably with the raw UBPC data (black dots) after amplitude scaling. This result shows that the UBPC EQE spectrum approximates well the joint density of states' stepwise functional form, and therefore, can be used as a proxy for absorption spectrum in the Tauc plot.

As a final integrity-check of our UBPC approach, in Fig. 3c EQE spectra are taken at temperatures increasing from 125 to 380 K. The blue and red dashed lines highlight the changing Urbach tail disorder, characterized by the Urbach energy, $E_U$, which increases from 94 to 128 meV for this temperature range. Similar trends are observed in other oxide semiconductors.[30, 31] The room temperature $E_U$ value for a-IGZO is found to be 111 meV. The inset in Fig. 3c shows the Urbach energy extracted from the EQE spectra taken at different temperatures. The dashed line in the inset of Fig. 3c represents the temperature-dependent Urbach energy, $E_U(T)$, calculated from[54] $E_U(T) = E_0 + E_1 \left(e^{\theta_E/T} - 1\right)^{-1}$ where $E_0$ is the temperature-independent structural disorder due to the lack of long-range order, $E_1$ parameterizes the temperature-driven disorder due to phonons, and $\theta_E$ is the Einstein temperature. We found that $E_0 \sim 80$ meV, $E_1 \sim 31$ meV and $\theta_E \sim 185$ K. The average energy of phonons in the lattice is then given as $k_B\theta_E = 15$ meV. This value is notably smaller than that of Si and Ge[55], and could lead to enhancement of electron-phonon scattering processes in a-IGZO at room temperature, compared to Si and Ge, since the phonon energy for a-IGZO is less than $k_BT$ (at room temperature).

**FIRST-PRINICPLE DoS ANALYSIS:**

In order to simulate donor- and acceptor-like states in the sub-gap of *a*-IGZO, two basic types of defects were considered: oxygen vacancies and metal vacancies. Each type of vacancy was simulated a multitude of times in order to get the energetic distributions of the donor and acceptor states. Further, considering the energy of defect formation for oxygen self-interstitials and metal vacancies, we find that the most likely candidate to explain the acceptor-like behavior shown in the Results is the Zn vacancy.

The energetic distribution of vacancy defects in *a*-IGZO was constructed from individual DFT+U simulations for 160 oxygen vacancies, and 60 metal vacancies. (See Supplemental Information for DFT+U simulation parameters.) A vacancy defect is created by removing either a neutral oxygen or a metal atom from a simulated *a*-IGZO cell to form either $V_O$ or $V_M$, respectively. Two stoichiometric *a*-IGZO cells, each consisting of 140 atoms, were used as test structures for the vacancy DoS analyses. For each simulation, one atom was removed, and the atomic structure was relaxed to minimize energy in the lattice. To compare the relative vacancy defect energies in reference to the conduction and valence bands, the energy axes of each simulation were scaled to approximate the experimental *a*-IGZO bandgap. A composite DoS is then constructed by taking the sum of all individual vacancy simulations.

Figure 4ai shows the DoS for a pristine *a*-IGZO cell (i.e. no vacancies). The Fermi-level is positioned to the right of the last filled state near the VB. In general, after an oxygen vacancy

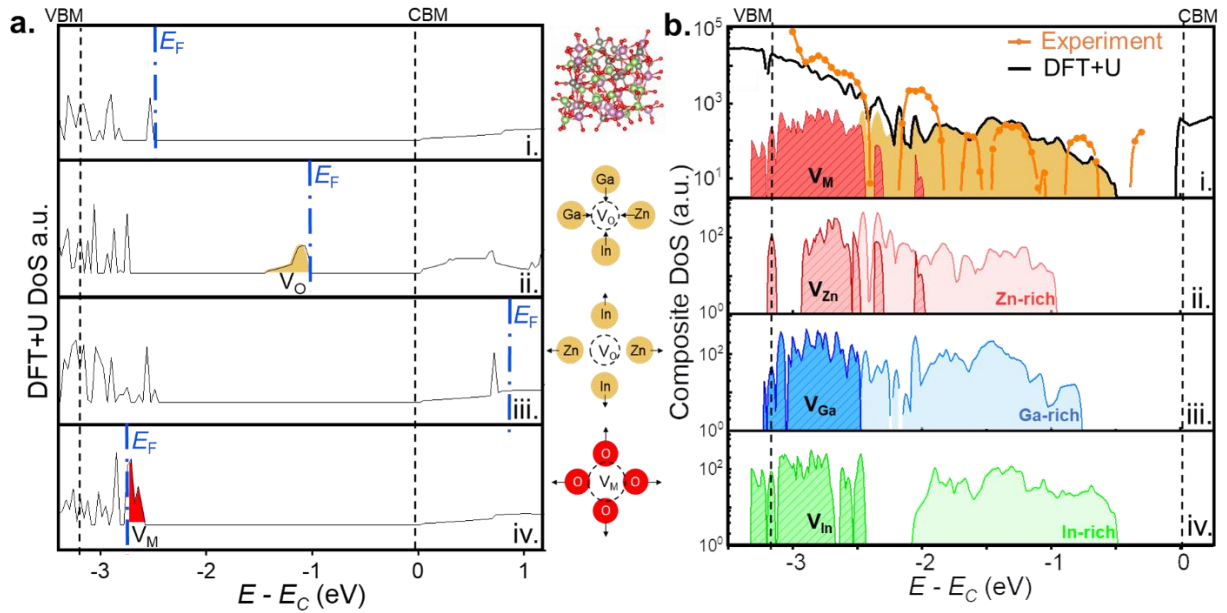

**Figure 4. (a)** DFT+U DoS simulations for a-IGZO. **i**. Pristine a-IGZO cell (*structure on right*) shows $E_F$ is located near the valence band. **ii**. An oxygen vacancy ($V_O$) creates a deep donor state (*shaded gold*) and $E_F$ shifts toward CB. **iii**. A different $V_O$ creates a shallow donor state and $E_F$ shifts into the conduction band. **iv**. A metal vacancy ($V_M$) creates a deep acceptor state (shaded red) and $E_F$ shifts away from CB. **(b) i.** The composite DoS is the sum of DFT+U vacancy simulations, with overlay of experimental data. Contribution of $V_M$ states is shaded red, and $V_O$ are shaded gold. **ii. - iv.** Coordination-specific composite DoS show the sum of last filled states due $V_O$ deep donors with majority In, Ga or Zn as nearest neighbors (In-, Ga-, or Zn-rich; *light shading*), and the sum of unfilled $V_M$ acceptor states due to In, Ga, Zn, removal (*hashed shading*)

is created (prior to structural relaxation), a deep state forms in the sub-gap. However, after the structural relaxation, either a "deep state" forms in the sub-gap due to the inward relaxation of the coordinating metals (e.g., Fig. 4aii, gold shading); or a "shallow state" is formed in the conduction band due to the outward relaxation of the coordinating metals (e.g., Fig. 4aiii). The formation of a deep state indicates that a trap for electrons is created in the sub-gap, whereas the formation of a shallow state indicates that electrons are donated to the CB. By contrast, the creation of a neutral metal vacancy tends to increase VB tail state disorder while the Fermi-level shifts further away from the CB, leaving unfilled states in the sub-gap (e.g. Fig. 4aiv, shaded red). The fact that neutral oxygen vacancies shift $E_F$ to the right of states, meaning they are filled, is a direct indication that oxygen vacancies are donor states. Conversely, neutral metal vacancies shift $E_F$ to the left of states, leaving them empty, which is a direct indication that metal vacancies are acceptor states.

Figure 4bi shows the composite DoS, which is the summation of all DoS obtained from individual vacancy defect simulations (black line). This is overlaid with the experimental DoS (orange line) for comparison. There is good agreement between the defect distributions suggested by the composite DoS and experiment, as both indicate defect states span, virtually, the entire sub-gap. It is necessary to compare the sum of all DoS from vacancy defect simulations (as opposed to an individual vacancy defect simulation) with the experimental DoS, since the experiment represents a stochastic spatial average of all defects contained within the laser excitation area. 220 vacancy defects are sufficient to provide a representative sample of metal-oxygen (M – O) coordination environments (see Supplemental Information). This is important since TFT behavior depends on the stochastic defect population, and so consideration of the multitude of defect states is required to accurately describe the physical system.

A clear indication of the energetic distribution of oxygen versus metal vacancies is shown in Fig. 4bi. The gold shaded area is the sum of the last filled states for all deep $V_O$ (e.g., Fig. 4aii), and the red shaded area is the sum of the unfilled states for each $V_M$ (e.g., Fig. 4aiv). Thus, according to DFT + U simulations, $V_M$ acceptor states are found from VBM to ~ −2 eV, and $V_O$ donor states are found from −2.5 eV to −0.5 eV. These sub-gap $V_O$ donor states account for 53% of oxygen vacancy simulations leading to the formation of "deep states," while the remaining 47% led to "shallow" states forming in the CB.

The distribution of deep sub-gap states arising due to different M – O coordination environments is shown in Figure 4bii–iv. Oxygen vacancies were classified by considering the metal atoms within 2.5 Å of the oxygen location, i.e. their nearest neighbors. Oxygen atoms had either three or four neighboring metal atoms. (See Supplemental Information for details of oxygen nearest neighbors.) $V_O$ deep donor sites with a majority of In atoms (In-rich) as nearest neighbors (Fig. 4bii, light green shading) tend to form defect states closest to the CBM. Deep $V_O$ sites having majority Ga atoms as nearest neighbors (Ga-rich) are shifted slightly further from away from the CB (Fig. 4biii, light blue shading). Deep $V_O$ sites with a majority of Zn atoms as nearest neighbors (Zn-rich) tend to form the deepest states on average (Fig. 4biv, light red shading).

Creation of a neutral zinc, gallium, or indium vacancy ($V_{Zn}^0$, $V_{Ga}^0$, $V_{In}^0$) gives rise to trap states positioned within the valence band tail state portion of the IGZO band gap, (Fig. 4bii, iii and iv, hashed red, blue and green shading,) with these states being primarily formed from O-2p basis functions. The indium and gallium vacancy trap distributions are continuous and very broad, i.e., ~0.8 and ~1 eV, respectively, while the main zinc vacancy peak is less broad, i.e., ~0.5 eV, and three zinc vacancy satellite peaks are present. The broad energetic spread of these distributions is attributed to the amorphous nature of *a*-IGZO. The energetic locations of the upper two zinc vacancy satellite peaks are in good agreement with UBPC trap states Zn-7 and Zn-8 of Table 1. This is strong supportive evidence for ascribing these peaks to zinc vacancies.

More evidence for our zinc vacancy identification is obtained by considering the energy of formation of vacancy defects in oxygen-poor *a*-IGZO. We find from DFT assessment that the energy of formation for each neutral vacancy type is as follows: neutral oxygen vacancy = -3.35 ± 0.86 eV; neutral zinc vacancy = 3.84 ± 0.82 eV; neutral indium vacancy = 6.18 ± 0.82 eV; neutral gallium vacancy = 8.40 ± 0.9 eV. However, this is not the complete story. Most oxygen vacancies are expected to be neutral while all three metal vacancies are expected to be ionized, since the position of the Fermi level is near typically near the conduction band mobility edge in *a*-IGZO.

Figure 5 shows the metal vacancy and oxygen self-interstitial defect formation energy as a function of Fermi-level in oxygen-poor *a*-IGZO. (See Supplemental Information for details on defect formation energy calculation.) The points on each line indicate ionization energies and are labeled corresponding to the vacancy charge when the Fermi level is below/above the ionation energy.

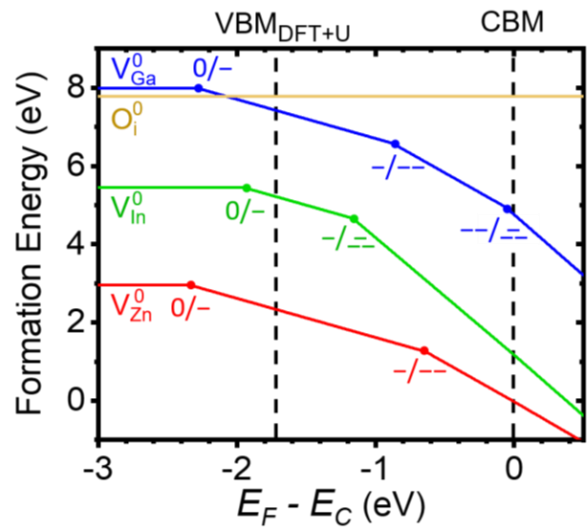

**Figure 5.** Formation energy versus Fermi level for gallium, indium, and zinc vacancy and oxygen self-interstitial acceptor-like defects using the DFT+U band gap for a-IGZO under oxygen-poor conditions.

These computations indicate that when the Fermi level is positioned near the conduction band mobility edge, a zinc vacancy has a very small formation energy compared to that of an indium vacancy or, especially, a gallium vacancy or oxygen self-interstitial. Moreover, the zinc vacancy formation energy is found to be very small, and sometimes negative such that zinc vacancy creation is energetically favorable as a consequence of self-compensation (see Discussion). Although formation energy versus Fermi level plots are shown for a zinc, indium, and gallium vacancy, and an oxygen self-interstitial, it is important to note that these plots are representative of general trends and that a substantive variability in the absolute values are expected due to the amorphous nature of *a*-IGZO. The ionization energy trends shown in Figure 5 suggest that zinc and gallium vacancies are positive-U defects, while an indium vacancy is a negative-U system with respect to the -/- - ionization state. The ionized oxygen self-interstitial states (1- and 2-) had higher defect formation energy than the neutral vacancy when the Fermi-level was contained in the band gap. Thus, the energetics of vacancy formation in a-IGZO support the notion that the peaks labeled Zn-7 and Zn-8 are indeed associated with the presence of zinc vacancies, and (possibly, but not likely) indium vacancies, but not gallium vacancies or oxygen self-interstitials.

**DISCUSSION:**

At first glance, the existence of zinc vacancies in *a*-IGZO is unexpected and even unsettling, since it is well known that *a*-IGZO (and almost all oxides) tend to be oxygen deficient. Why would cation vacancies form in a material that is anion deficient? The answer to this puzzle is self-compensation, as discussed below.

Since *a*-IGZO is oxygen deficient, it tends to have a large density of oxygen vacancies. Our DFT computations indicate that oxygen vacancies are donors. Moreover, we find that some of these oxygen vacancies are deep in the gap, and therefore neutral, while other oxygen vacancies are shallow, and hence ionized. The existence of ionized donors pushes the Fermi level towards the conduction band mobility edge. Typically, the Fermi level in *a*-IGZO is positioned within about 0.2-0.3 eV from the conduction band mobility edge. This positioning of the Fermi level near to the conduction band mobility edge strongly reduces the formation energy of an ionized zinc vacancy compared to what it would be if it were neutral (Figure 5). When the Fermi level is positioned close enough to the conduction band mobility edge that the formation energy of the zinc vacancy becomes negative, it is more energetically favorable for the next donor atom to energetically relax by ionizing a newly created zinc vacancy (compensation) than to simply promote the electron to the conduction band in the normal manner. Note that self-compensation by zinc vacancy creation only occurs in the presence of a much larger concentration of oxygen vacancies such that *a*-IGZO is indeed oxygen deficient, as expected.

The existence of zinc vacancies in *a*-IGZO appears to be deleterious from the perspective of TFT operation since photo-ionization of these states is likely responsible for the negative bias illumination stress (NBIS) instability. If zinc vacancies are indeed the culprit responsible for NBIS, it might be worth exploring *a*-IGO as an alternative TFT channel layer material since indium and gallium vacancies have much larger formation energies.

**CONCLUSIONS:**

In summary, the ultrabroadband DoS of *a*-IGZO is measured spanning the sub-gap states (0.3 to 3.5 eV) using our ultrabroadband photoconduction (UBPC) technique. Eight sub-gap peaks are correlated to the local coordination environments of vacancy defects using density functional theory, and transient PC measurements. UBPC measurements give a Tauc-gap (~ 3.15 eV) and VB Urbach energy (110-120 meV, at room temperature) across multiple *a*-IGZO TFTs. Six donor-like oxygen vacancy peaks dominate the majority of the sub-gap extending to ~ 2.5 eV below the conduction band edge. Two acceptor-like Zn vacancy peaks are also detected at about 2.1 eV and 2.6 eV below the conduction band edge. *Ab initio* calculations of DoS and formation energy identify these experimentally observed acceptor-like peaks as predominantly zinc and, possibly, indium vacancies. TFT characteristics such as threshold voltage and photoinduced hysteresis are shown to be directly related to the density of Zn vacancy states. The impact of the Zn vacancy concentration on TFT threshold voltage is a consequence of detailed charge balance in *a*-IGZO, given their acceptor-like nature. In addition, electron recombination is found to be slower when electrons are photoexcited from Zn vacancy states than from oxygen vacancy states. This trend is ascribed to the acceptor- or donor-like nature of the Zn or oxygen vacancy, respectively, due to neutral or coulombically attractive capture. By identifying the energy distributions and, moreover, the electronic configurations of vacancy defects through density functional theory, this work reveals the unanticipated impact Zn vacancy deep acceptors play in tuning the electronic properties of *a*-IGZO TFTs.

**Corresponding Authors**
* kyletvogt@gmail.com; paulc@science.oregonstate.edu;

 jfw@eecs.oregonstate.edu; graham@physics.oregonstate.edu

**PRESENT ADDRESS**
†Department of Physics, Broida Hall, University of California, Santa Barbara, Santa Barbara, CA 93106-9530

# Supplementary Materials:

## Ultrabroadband Density of States of Amorphous In-Ga-Zn-O

K. T. Vogt[1]*, C. E. Malmberg[2], J. C. Buchanan[2], G. W. Mattson[1],
G. M. Brandt[1] D. B. Fast[2], P. H.-Y. Cheong[2],* J. F. Wager[3],* M. W. Graham[1]*

[1.] Department of Physics, Oregon State University, Corvallis, OR 97331-6507 USA
[2.] Department of Chemistry, Oregon State University, Corvallis, OR 97331-4003 USA
[3.] School of EECS, Oregon State University, Corvallis, OR 97331-5501 USA

## S.1: Figure 1 Device Transfer Curve

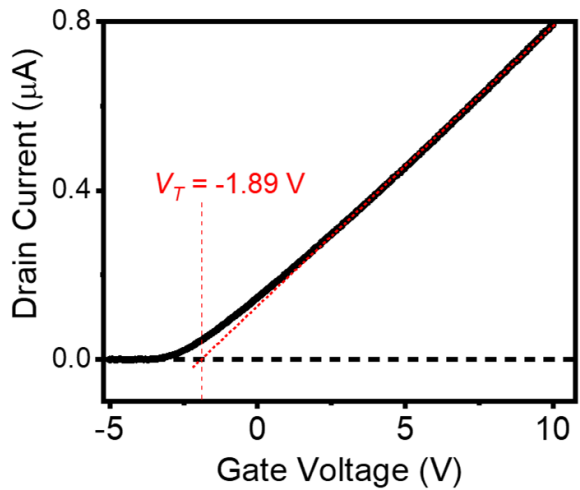

$I_D - V_G$ transfer curves from device measured shown in Figure 1 shows a threshold voltage of $V_T \sim -1.89$ V in Fig. S1, corresponding to the value predicted for out experimentally measured deep acceptor trap density (see Fig. 1c, red dashed line).

## S.2: UBPC Experimental Setup Details

UBPC experiments are performed using tunable lasers coupled into two modified home-built scanning photocurrent confocal microscopes (SPCM) that yielded consistent $a$-IGZO DoS reconstructions. SPCM microscope #1 used an ultrafast Ti:Sapphire laser (Coherent Chameleon Ultra II) pumping an Optical Parametric Oscillator (APE Compact) to generate a continuous $hv$ = 0.3 to 3.5 eV tuning range. SPCM microscope #2 used a long-pulse Fianium SC400 supercontinuum white light laser, from which photon energies ($hv$ = 0.7 to 3.1 eV) are scanned using a laser line tunable filter (LLTF, Photon Etc.). Various longpass filters (>OD6) prevent higher-order leak contributions to UBPC. All optical elements including homebuilt reflective SPCM optics and a piezo-scanning mirror were designed to maintain beam Poynting stability at <1 μm on $a$-IGZO. Typical illumination beam fluxes are ~$10^{13}$ photons/cm$^2$ and illumination spot-size is measured to be near the diffraction limit. Both setups used an Olympus BX-51WI microscope with a 52X TechSpec Adjustable ReflX Objective (Edmund Optics). The UBPC data is collected by $rf$ micromanipulator-probes connected to Yokogawa GS200 DC source meter. After current pre-amplification (Ithaco 1211), the UBPC signal is isolated by lock-in amplifier (Zurich HF2LI) phase-locked to an optical chopper. High-sensitivity calibrated power meters and bolometer (Thorlabs, PM16-401 and S120VC) are used to normalize for the incident power after the objective.

Carrier lifetime traces were captured using a Tektronix TDS 3054B 500 MHz oscilloscope (~700 ps rise time) while being synced to the chopping frequency of ~240 Hz. The data presented in the paper was from 32 consecutive averages of the photocurrent decay times. Figure 2b includes the composite instrument response (gray dotted line), which was limited by the time it took for the chopper to traverse the laser spot. The chopper intersected the laser at the focal point of a pair of identical 5 cm lenses. The Ithaco 1211 current pre-amplifier was set at $10^{-6}$ A/V sensitivity, with a time constant of 1 ns. From this, it is clear that the hardware is completely capable of measuring the fastest response we observed (the instrument response) and therefore can be trusted to accurately represent the decay time of photoexcited electrons when searching for the slowest time constants

Temperature-dependent measurement with conducted over a 77 to 360 K range in a temperature-controlled optical cryostat (Advanced Research Systems, LN2 continuous flow).

## S.3: DFT Simulation Parameters Details

For DFT + U simulations, amorphous structure cells were first generated by placing twenty formula units of IGZO at random into a cubic lattice cell with 30% more volume than the crystalline structure. The cells were then taken through melt and quench molecular dynamic simulation via the General Lattice Utility Program (GULP)[1]. Specifically, the cells were melted to 4000 K for 30 ps and then cooled by 100 K/ps until the system reached 300 K, the cell was then equilibrated at 300 K for 3 ps[2]. The resulting structures were then refined using Vienna ab-initio Simulation Package (VASP). The projector augmented wave (PAW) method was employed using the Perdew, Becke, and Ernzerhof (PBE) Generalized Gradient Approximation (GGA) functional[3–7] theory. Three individual relaxations were done using a 1x1x1 k-point mesh centered at the gamma point, the conjugate gradient algorithm was used. The first and third relaxations only allowed the ions to relax, while the second relaxations only allowed for the cell sizes to change. This resulted in amorphous IGZO structures with a density of ~6.1 g/cm$^3$, which is in agreement with results reported by Kamiya et al.[8] All density of states spectra were computed using GGA+U with a 4x4x4 Monkhurst pack k-point mesh. The U-values employed for the In 4d, Ga 3d, Zn 3d, and O 2p were 7, 8, 8, and 7 respectively, the In/Ga/Zn values were extracted from Noh, Chang, Ryu, and Lee[9] and the value for oxygen was taken from Ma, Wu, Lv, and Zhu[10].

Formation energies of all defects were calculated by using the Freysoldt-Van de Walle method[11] using equation 1:

$$E^f[X^q] = E_{tot}[X^q] - E_{tot}[bulk] - \sum n_i \mu_i + qE_F + E_{corr} \tag{1}$$

Where $E_{tot}[X^q]$ is the total energy of the defect containing super cell, $E_{tot}[bulk]$ is the total energy of the defect-free bulk super cell, $n_i$ is the number of atoms added or removed of a given element $i$ (where $i$ = In, Ga, Zn, O) to create the defect structure, $\mu_i$ is the chemical potential of element $i$, $E_F$ is the fermi energy in reference to the conduction band minimum, and $E_{corr}$ is a correction term for the total energy of the charged defect that accounts for the electrostatic interactions between the periodic images of the supercell due to their finite-size.[11,12]

The chemical potentials of each atom were calculated as outlined in S. KC et al.,[13] where each chemical potential for a given element $i$ was defined as:

$$\mu_i = E_i + \Delta\mu_i \tag{2}$$

where $E_i$ is the total energy of element $i$ in its elemental phase, and $\Delta\mu_i$ is the change in the chemical potential depending on the growth and annealing conditions to be examined. To make sure that a realistic range of values for $\Delta\mu_i$ are chosen for each element in a-IGZO, the thermodynamic stability window of a-IGZO is required. Using the following equations

$$\Delta\mu_{In} + \Delta\mu_{Ga} + \Delta\mu_{Zn} + 4\Delta\mu_O = H^f(a\text{-IGZO}) \tag{3}$$

$$\Delta\mu_{In}, \Delta\mu_{Ga}, \Delta\mu_{Zn}, \Delta\mu_O \leq 0 \tag{4}$$

$$\Delta\mu_{Zn} + \Delta\mu_O = H^f(\text{ZnO}) \tag{5}$$

$$2\Delta\mu_{Ga} + 3\Delta\mu_O = H^f(\text{Ga}_2\text{O}_3) \tag{6}$$

$$2\Delta\mu_{In} + 3\Delta\mu_O = H^f(\text{In}_2\text{O}_3) \tag{7}$$

$$\Delta\mu_{In} + \Delta\mu_{Ga} + 3\Delta\mu_O = H^f(\text{InGaO}_3) \tag{8}$$

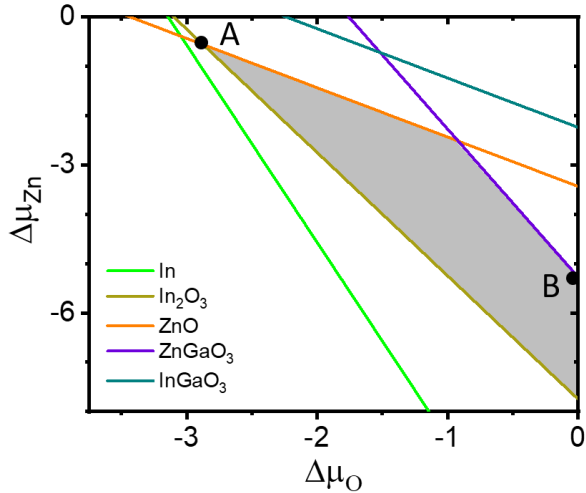

we can plot the stability region of a-IGZO. Figure S2 shows the stability region of a-IGZO with respect to the $\Delta\mu_{Zn} - \Delta\mu_O$ phase space, from this plot the oxygen-poor (point A) and oxygen-rich (point B) chemical potentials can be derived for each atom.

**Figure S2:** Chemical potentials with respect to the $\Delta\mu_{Zn} - \Delta\mu_O$ phase space. The stability window is shaded in gray. Points A and B correspond to O-poor and O-rich chemical potentials, respectively.

## S.4: Metal – Oxygen Coordination Environments

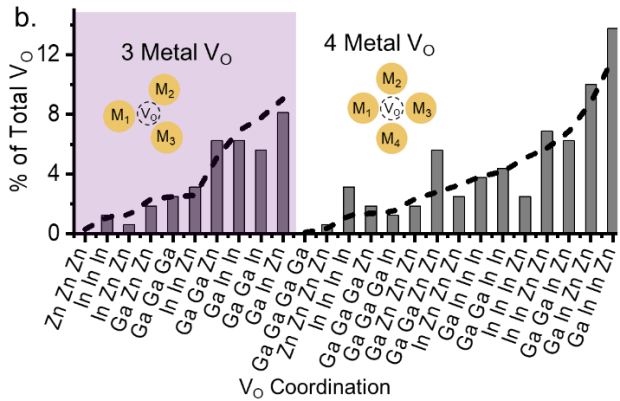

**Figure S3:** Coordination environments for oxygen vacancies separated by those surrounded by three metals (purple shaded area) and four metals (no shading). The gray bars indicate the frequency (%) of each M-O coordination in this work. The black dashed is the expected frequency (%) of each M-O coordination in the stochastic limit

Figure S3 summarizes the nearest atom neighbors within 2.5 Å for the 160 oxygens in the amorphous cells. The overwhelming majority of oxygen atoms possessed either three or four metals as nearest neighbors (~ >99%).

In order to show that our sample was large enough to provide meaningful statistics for the coordination environments, we compared the computed distribution of oxygen environments from this work to the ergodic distribution previously reported by our group.[2] The ergodic distribution consisted of 55 cells, each with 240 atoms, thus a total of 13,200 metal-oxygen coordination environments; these are shown in dashed line in Figure S2. This comparison reveals that the 160 oxygen vacancies show significant signs of convergence to the ergodic limit.

In the case of metal vacancies, only 60 metal atoms were considered, as the nearest neighbors of $V_M$ are all O atoms and virtually homogenous.

## S.4: References